# Scalable platform enabling reservoir computing with nanoporous oxide memristors for image recognition and time series prediction

*Joshua Donald, Ben A. Johnson, Amir Mehrnejat, Alex Gabbitas, Arthur G. T. Coveney, Alexander Balanov, Sergey Savel'ev, Pavel Borisov\**

J. Donald, B. A. Johnson, A. Mehrnejat, A. Gabbitas, A. G. T. Coveney, A. G. Balanov, S. Savel'ev, P. Borisov
Department of Physics, Loughborough University, Loughborough, LE11 3TU, United Kingdom
E-mail: p.borisov@lboro.ac.uk
A. Mehrnejat
Department of Physics & Astronomy, University of Manchester, Manchester, M13 9PL, United Kingdom

Funding:
UK Engineering and Physical Sciences Research Council (EPSRC) studentship 2764909
UK Engineering and Physical Sciences Research Council (EPSRC) research grant EP/T027479/1

Abstract
**Typical mammal brains have some form of random connectivity between neurons. Reservoir computing, a neural network approach, uses random weights within its processing layer along with built-in recurrent connections and short-term, fading memory, and is shown to be time and training efficient in processing spatiotemporal signals. Here we prepared a niobium oxide-based thin film memristor device with intrinsic structural inhomogeneity in the form of random nanopores and performed computational tasks of XOR operations, image recognition, and time series prediction and reconstruction. For the latter task we chose a complex three-dimensional chaotic Lorenz-63 time series. By**

**applying three temporal voltage waveforms individually across the device and training the readout layer with electrical current signals from a three-output physical reservoir, we achieved satisfactory prediction and reconstruction accuracy in comparison to the case of no reservoir. This work highlights the potential for scalable, on-chip devices using all-oxide reservoir systems, paving the way for energy-efficient neuromorphic electronics dealing with time signals.**

## 1. Introduction

The current development of neural networks and artificial intelligence (AI) in general is still deficient in terms of the energy efficiency of the hardware used to implement it.[1–5] At the same time, the efforts to train the relevant networks, in particular the ones which handle time series (language recognition and sensor data processing for example) are still considerable. When operating, those networks often rely on maintaining connections to online servers which makes their operation much less secure, slower and energy inefficient. In this context, the approach of reservoir computing realized in physical hardware is promising in terms of creating compact, energy-efficient devices which can be easily trained and operated in offline mode.

Reservoir computing (RC) is a subset of recurrent neural networks.[6–8] But also, RC has some origins in the computational models of interactions between the primate cortex and the striatum in the brain.[9] The hidden layer, the so-called reservoir, features random weights that are fixed after initialization whilst only the output layer weights are trained. The built-in recurrent connections in the reservoir allow it to deal with time series data due to the corresponding short-term, fading memory and increased dimensionality. The training can happen at much lower costs, by a simple linear regression algorithm, as it is limited to the output layer only. The non-linearity of the reservoir as well as its higher dimensionality, compared to that of the input, allows one to classify the input signal or predict the future time series evolution. All these properties allow particular efficiency in processing temporal data with a simplified training process.[10]

Currently, RC is widely used for prediction and recognition tasks related to complex dynamics time series, such as: mathematical chaos,[11] environmental sensors,[12] stock market indices,[13] robotic arm control,[14] fluid dynamics,[15] weather,[16] climate change[17] and heart and brain medical diagnostics.[18]

The fact that the reservoir can be left unmodified with random initial weights makes it relatively easy to use physical systems and devices as physical reservoir realizations with the promises of offline operation (edge computing), low latency, and much better energy efficiency.[19,20]

Examples include magnetic[21,22] and photonic systems[23–25] and oxide memristors.[26–36] Typical tests included pattern recognition,[25–28,30,31,33,36] time series prediction[22,24,29,32–36] and nonlinear dynamics reconstruction.[28,30,34]

Volatile oxide memristors, for example those with $NbO_x$ switching layers,[37–41] are two-terminal devices which switch to a low resistive state (LRS) when the applied voltage is above a threshold and return to a high resistive state (HRS), with some hysteresis, upon removal of bias voltage. Energy efficiency, scalability and compatibility with standard computational hardware are some examples of promising and desirable features of volatile oxide memristors. These are suitable candidates to be applied to physical RC due to their ability to store information as short-term memory (hysteresis),[42–44] directly comparable to the process of short-term plasticity (STP) observed in biological systems.[45] Memristor reservoirs were created either using single or multiple devices[26,30,33] or in a crossbar array configuration.[28,29,33,36] This approach brings with it the issues of poor scalability, limited device interconnections and structural complexity. Jaafar et al. used mesoporous $SiO_2$[31] layers in order to enhance the diffusivity of Ag ions in the corresponding memristor devices and performed an image recognition task via the RC approach as a test. Here, from the start we intentionally create a thin oxide film device with random nanopores associated with different output channels, in order to embrace the randomness of such a reservoir, to allow the formation of multiple pathways for charge carriers to take, and thus, to enhance the capabilities of the reservoir for solving temporal tasks.

To better benchmark a reservoir, chaotic time series are often used as they are very challenging to forecast, as minor changes in their initial state generate very different future behaviors. The Mackey-Glass time series exhibits aperiodic behavior which can be difficult to predict for many machine learning systems, most notably the peaks of waveforms,[46] yet physical, memristor-based RC has proven capable of doing so.[29,32,36] This specific example of the Mackey-Glass time series operates through a single temporal waveform, defined entirely by one delay-differential equation.[47] Other sought after chaotic systems exist consisting of multiple differential equations, such as the Lorenz system, governed by three equations and thus possesses three dimensions.[48] The forecasting of the Lorenz model has also been previously investigated using physical hardware in form of a photonic system[24] or a single memristor device with dynamic signal pre-processing and virtual reservoir outputs,[34] both finding relative success in reproducing or predicting the Lorenz time series. However, these forecasts did not delve into how well the system would have performed without a reservoir or performed at different input and output data set sizes. This is important to compare to benchmark the reservoir's performance.

In this work, we demonstrate how a scalable platform of NbO$_x$-based volatile memristors with an intrinsic structural inhomogeneity due to nanoporosity can be used in physical RC systems and are capable of complex computations, with future potential for industry-accessible, on-chip devices, for example, predicting industry-relevant time series or complex atmospheric flows. The nanoporous morphology of the device allows the formation of multiple, randomly distributed pathways which enhance the temporal capabilities of the reservoir and mimic, in some way, the random network of neurons in the brain.

We test our physical reservoir's ability for performing the logical XOR operation and image recognition tasks and to predict and reconstruct the chaotic Lorenz system time series *X(t)*, *Y(t)* and *Z(t)*. We achieve the latter by applying these three temporal waveforms, individually, to the reservoir. By training the readout layer of our three-output reservoir, the temporal data of the chaotic system can be predicted and reconstructed with high accuracy.

## 2. Results and discussion
### 2.1. Physical reservoir device with intrinsic nanoporosity

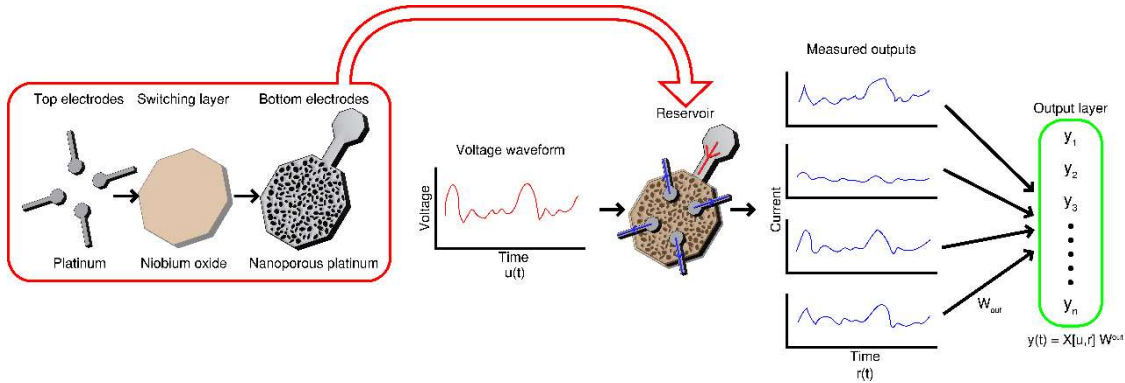

**Figure 1.** The principle of physical RC using a nanoporous memristor oxide device. From the left to the right: Composition of the physical reservoir, including the top electrodes, the memristor switching layer and bottom nanoporous layer. Schematic representation of the voltage input *u(t)*, current waveforms representing the reservoir states *r(t)*, output layer *y(t)* produced by a product of the concatenated input vector *X*[*u(t)*, *r(t)*] with the weight matrix $W^{\text{out}}$.

Figure 1 shows the composition of our nanoporous physical RC system and its principles of operation. The higher dimensionality and separability of the reservoir is assured by the introduction of nanoporous structures in the oxide switching layer, in order to design multiple devices capable of outputting varying readout values. This is realized through the template of a nanoporous Pt bottom electrode which is first deposited and annealed to create a nanoporous layer, before patterning via photolithography. The non-linearity and fading memory necessary for RC are then established through a resistive switching layer of nitrogen-doped $NbO_x$. This was deposited atop this nanoporous layer and sandwiched under a conventional Pt top electrode with a thin Ti intermediary adhesion layer. Resistive switching in our $NbO_x$ is volatile and operates through the process of Joule heating.[38] The resistive switching layer was doped with nitrogen for better and more consistent switching performance.[49] Due to the presence of nanopores in the bottom Pt electrode, the oxide layer possesses a similar, but not identical, structural inhomogeneity, meaning that each device's output on the chip is not identical to the others with respect to its local *I-V* response. Each reservoir output thus encompasses a unique combination of memristive conducting channels, formed by the corresponding pattern of nanoporosity present within that specific top electrode area. This structure is not dissimilar to the complex neural connections of a mammal brain.

The input time series were applied as a voltage waveform through the bottom electrode and up to four current waveforms were recorded simultaneously through the available reservoir output channels in the form of top electrodes. Those served as reservoir states with intrinsic recurrent connections (highly inhomogeneous current routes) due to the fading resistive switching memory and mimicking random resistance-based "weights" aided by the random distribution of current paths in the nanoporous layer and metal oxide film. The output was constructed by multiplying the concatenated vector $X[u(t), r(t)]$ of input $u(t)$ (voltage) and reservoir $r(t)$ (current data) time series with the weight matrix $\mathbf{W^{out}}$ in an external program code.

## 2.2. Nanoporous memristor device

A top view of one of our nanoporous devices for RC is shown in the scanning electron microscopy (SEM) image in Figure 2A. A nitrogen-doped $NbO_x$ thin film with 80 nm thickness was sandwiched between four separate 120 nm top Pt electrodes and a common 20 nm nanoporous bottom Pt electrode (see Methods for details). The top electrode width dimensions were 10 μm. The SEM image for the nanoporous Pt is shown in Figure 2B, with the typical pore radius of $29 \pm 11$ nm, calculated using the method from.[50,51] The cross-section view in Figure 2C, acquired using focused ion beam (FIB) milling, shows the main layers under a single top Pt electrode. Variations in the brightness of the $NbO_x$ cross-section demonstrate nanoporosity, this being caused by the bottom np-Pt electrode. Figure 2D and Figure 2E show the chemical contrast in SEM images with Energy Dispersive X-ray (EDX). Analysis of the cross-section allows us to see all distinct layers (Pt, $NbO_x$ and Ti) as well as pores in Pt as regions that appear darker.

A single pore is observed in Figure 2E through higher resolution imaging of a thinner region of the FIB-cut cross-section. Comparing the EDX images of this isolated pore clearly illustrates how the deposited NbO$_x$ film does indeed fill the pore structure.

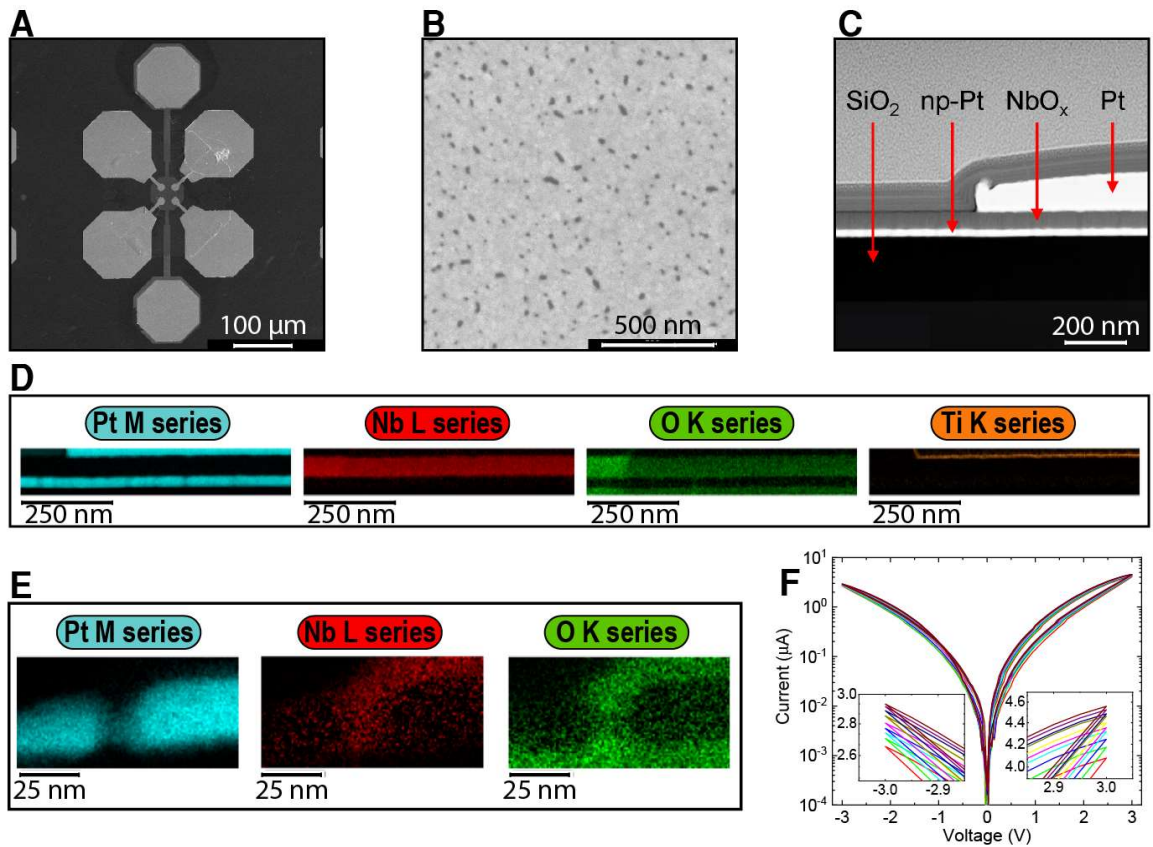

**Figure 2.** (A): SEM image of the device comprising of nanoporous (np)-Pt/nitrogen-doped NbO$_x$/Ti/Pt, with a np-Pt bottom electrode (20 nm), nitrogen-doped NbO$_x$ switching layer (53 nm) and four top Pt electrodes (120 nm). (B): Top view SEM image of the np-Pt bottom electrode; (C), (D), (E) cross-section images of the device from (A): general SEM (C), SEM with EDX analysis (D) for Pt (light blue), Nb (red), oxygen (green) and Ti (orange); magnified view of SEM EDX images (E) of a different cross-section where a nanopore in Pt (light blue) bottom electrode can be seen as filled by both Nb (red) and oxygen (green) denoting NbO$_x$ (F): Consecutive *I-V* sweeps of a singular nitrogen-doped NbO$_x$ based output channel shown in (A). Insets highlight current behavior between sweeps.

To study the resistance switching, voltage was swept from 0 V to 3 V and then back to 0 V before being repeated in the negative range with voltages of the same magnitude. The bottom electrode was kept at the ground potential. The change in resistance is continuous from HRS to the LRS, i.e. not abrupt (analogue switching), and is reversed back to the HRS once the external voltage is removed. Thus, a volatile resistive switching is observed in both voltage polarities, with hysteresis in the current-voltage loops. Figure 2F with insets shows 10 consecutive *I-V*s and the effect of these consecutive voltage sweeps. The maximal current values increase and decrease by 1-2% at each voltage sweep cycle in the positive and negative polarity, respectively. Hence, we deem the repeatability as sufficient for computing properties. Different reservoir output channels (top electrodes) of the same device demonstrated qualitatively similar yet quantitatively distinct *I-V* responses as desired for RC computations. The only exception was the unreliable fourth channel, which would typically be too conductive and provide a linear current response. Therefore, subsequent computations were performed on the three other channels only.

### 2.3. XOR task

To perform the first initial test of the RC performance of our devices we investigated the XOR task, that is, the ability of the system to compute an XOR operation by transforming the logical binary input pairs (0,1) or (1,0) to the output (1), and (0,0) or (1,1) inputs to the output (0). Despite its simplicity, the XOR task cannot be solved by a single layer perceptron (SLP).[52] Given that the output layer of our reservoir is indeed a SLP, if the reservoir layer is omitted, this is a useful test as it should fail if the reservoir does not provide any benefits to the computational performance.

To verify the SLP performance alone is insufficient, we used a standard SLP model provided by MATLAB R2024a, with a Heaviside activation function. It was used to classify the four input pairs of XOR tasks, and as a result, the training of the network failed, achieving 50% accuracy equivalent to a random guess between 1 or 0.

To construct a physical RC, we applied three consecutive voltage pulses, with the top electrode at the ground potential, where the first two pulses with 3 ms time duration were representing logical "1" or logical "0" with 1.75V or 0V amplitudes, respectively, while the last 1.75V pulse with 1.5 ms duration was used to read the resulting current as a corresponding reservoir output. A negative voltage was used to determine if the device would provide useful outputs in both the positive and negative voltage regimes as interpreted from the bipolar *I-V*s (Figure 2F.). Both voltages and currents produced by the four combinations of logical "0" and "1" inputs are shown in Figure 3A. Currents were recorded simultaneously from the three output channels in parallel (i.e. from three physical nodes for the RC outputs), but six values were taken from the readout pulse as virtual nodes. This provided more data to train and test with, which was then normalized as Z-scores for future calculations (see Methods for details).

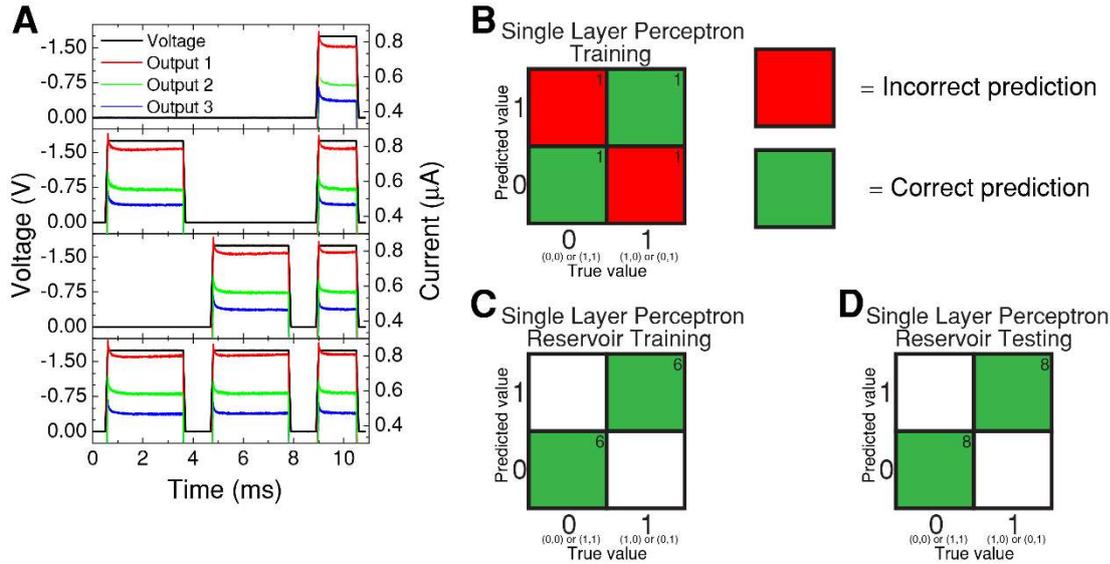

**Figure 3.** A): The four distinct pulse sequences used to produce read currents from three output channels simultaneously as corresponding to the XOR task inputs; Confusion matrices for the XOR task obtained during (B) training on SLP alone; (C) training and (D) testing on the same SLP using reservoir current outputs to the input voltage pulses from (A) with 12 and 16 data entries, respectively.

Contrary to the standard SLP model, our physical RC successfully realized the XOR task classification of the Boolean input pairs when the same SLP model was used on Z-scores calculated from the current output of the reservoir. See Figures 3B, 3C and 3D for training and testing confusion matrices, respectively. The training required only 12 iterations to achieve 100% accuracy, with the same performance observed from the test data.

This relatively simple XOR task is used as a first experimental test for our reservoir's ability to project an input signal into a higher dimensional output space. As intended, the three outputs in use provide a unique result. This higher dimensionality allows us to linearly separate reservoir outputs in contrast to a SLP alone.

### 2.4. Number recognition task

A further task performed by our RC system was recognition of binary images 5 x 3 pixels of zero through nine digits (Figure 4A), designed similarly to Du et al.[28] Each of the five rows, made up of three pixels, in every image were converted into a sequence of voltage pulses. A red or white pixel corresponded to a 1.5V or 0V pulse of a 3.2 ms time duration, followed by a final, shorter readout pulse at the end of each sequence (see Figure 4B for digit "0", as an example). These five pulse sequences were applied across a single output channel (i.e. single physical node) for each digit and the resulting current response was measured in the form of five readout values per readout pulse (i.e. five virtual nodes, taken data every 0.05 ms). This resulted in a total of 25 current values for each digit, see Figure 4C for output current combinations produced for a single virtual node. The obtained data was divided into two datasets, 60% for training and 40% for testing, and then processed by a SLP as part of a RC network. The training itself employed logistic regression, see Methods for details.

It was found that 100% recognition could be achieved after 14 iterations of training (see confusion matrices in Figure S1), even though just a single output channel of the reservoir was used. Reducing the number of allowed iterations for the training lowered the accuracy of recognition. For example, after 10 iterations of training, in one testing set a 5 was also classified as a 3. Overall, digits 3 and 5 had a normalized certainty as low as 43% and 45% respectively for 10 iterations, possibly due to the difference in just one pixel between both images. With 14 iterations, the highest normalized certainty was seen for recognition of digits 1 and 7, obtaining upwards of 83% and 91% respectively.

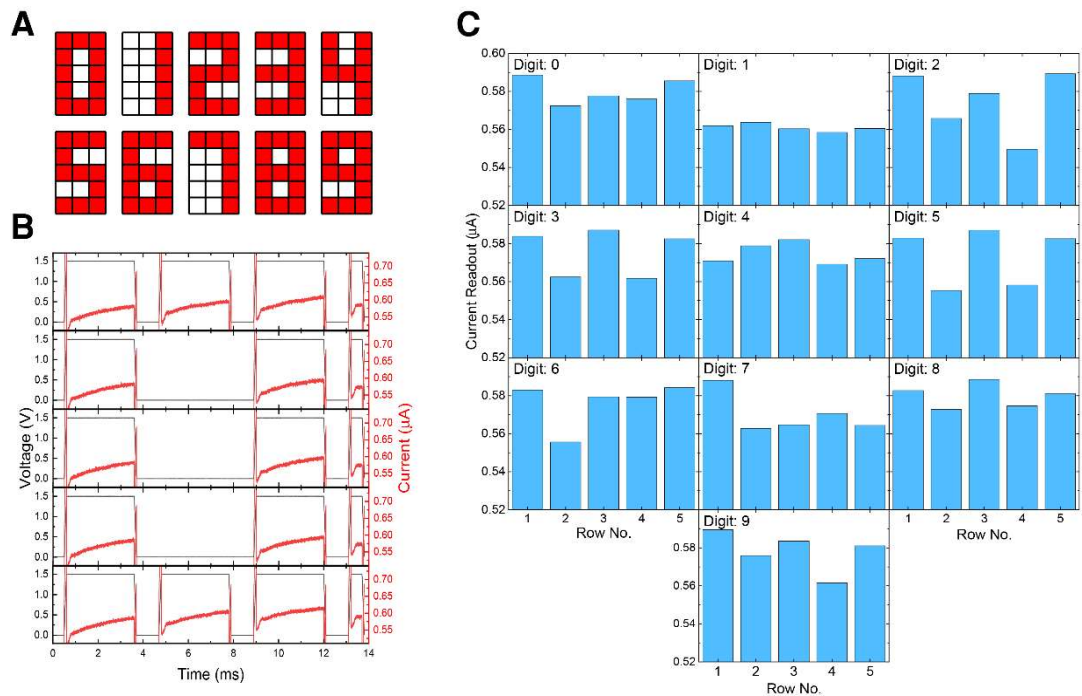

**Figure 4.** A): Binary digit numbers, zero through to nine; B): Voltage pulses (black line) with the corresponding currents (red line) applied to encode the binary image of digit "0", with five rows representing three pixels in each binary image row, followed by a shorter readout pulse. C): Current value combinations corresponding to ten binary digit images, as the reservoir output after five voltage pulse sequences were applied at the reservoir input for each digit image.

One current value from the readout pulse at the end of each voltage pulses sequence is plotted in each of the histograms.

## 2.5. Lorenz-63 system prediction

The most challenging computational tasks for our physical RC system are centered around the three-dimensional chaotic Lorenz time series {$X(t)$, $Y(t)$, $Z(t)$} described by the following differential equations and sometimes referred in literature as the Lorenz-63 model:[53]

$$\frac{dX}{dt} = \sigma(Y - X) \tag{1}$$

$$\frac{dY}{dt} = X(\rho - Z) - Y \tag{2}$$

$$\frac{dZ}{dt} = XY - \beta Z \tag{3}$$

In our calculations, we used the parameters: $s = 10$, $b = 8/3$ and $r = 28$[48] corresponding to the chaotic regime of the system. By solving the equations (see Methods for details) we obtained the time dependence for the three coordinates $X(t)$, $Y(t)$ and $Z(t)$.

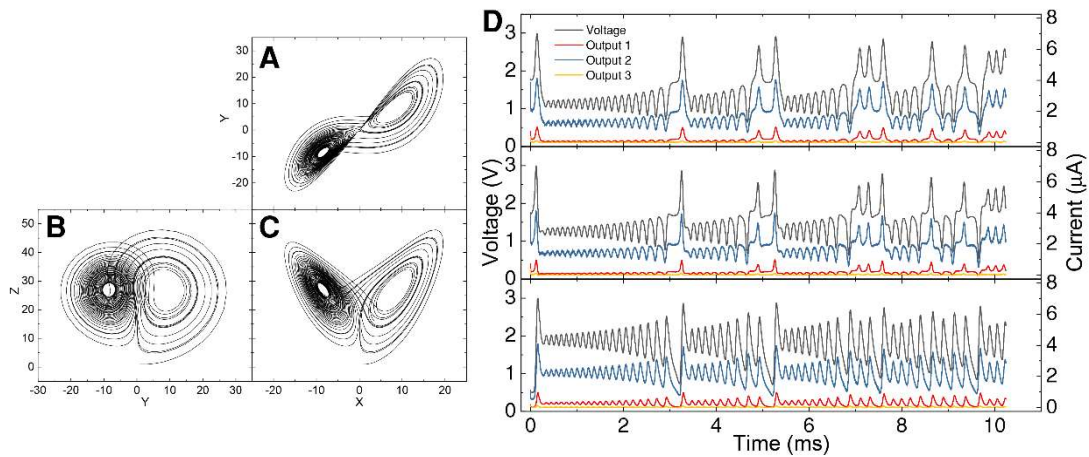

**Figure 5.** Lorenz chaos system for $\sigma = 10$, $\beta = 8/3$ and $\rho = 28$. Trajectory plots in the planes (A) $XY$, (B) $YZ$ and (C) $XZ$; D) Time series of three currents measured at each memristor device outputs when the input was driven by the voltage waveforms corresponding to $X(t)$, $Y(t)$ and $Z(t)$, from top to bottom.

Figure 5 (A-C) presents the projections of the corresponding Lorenz attractor onto the phase planes (*XY*), (*YZ*) and (*XZ*), respectively. They demonstrate chaotic orbiting around either of two unstable equilibrium points intermitted by occasional jumps between two of their vicinities, which reflects highly erratic nature of *X*(*t*), *Y*(*t*), and *Z*(*t*). The time series [*f*(*t*)=*X*(*t*), *Y*(*t*) or *Z*(*t*)] were then converted into voltage waveforms $V_X(t)$, $V_Y(t)$ and $V_Z(t)$, with an added voltage bias $V_0$ in order to keep the voltage in the positive range (see Methods for details). Here we present our results for the conversion parameters ($V_0$, $V_{Range}$) = (0.5 V, 2.5 V) where $V_{Range}$ is the voltage range, see discussion of other results at the end of this section. Figure 5D shows the corresponding voltages $V_X(t)$, $V_Y(t)$ and $V_Z(t)$ which were applied separately and sequentially to the common reservoir input (bottom electrode, Figure 1) together with the currents $I_{X1,2,3}(t)$, $I_{Y1,2,3}(t)$ and $I_{Z1,2,3}(t)$ which were then recorded at the three operating output channels simultaneously (see Figure 1). As can be seen in Figure 5D, each channel output realized a unique, non-linear transformation of the original input with the corresponding intrinsic recurrent (short-term memory) connections. We then constructed and trained a readout layer to predict the voltages in each of the time series $V_X(t)$, $V_Y(t)$ and $V_Z(t)$ for a certain number of time steps in the future ("future window") based on voltage (reservoir input) and currents (reservoir output) data points taken for a certain number of time steps in the past ("past window"). Both the voltage and current times series were normalized (see Methods for details). The prediction algorithm is an open loop, that is, a true signal data is used as an input for a given past window to predict new data in the future, and the process is re-iterated by shifting the prediction window (Figure 6A).

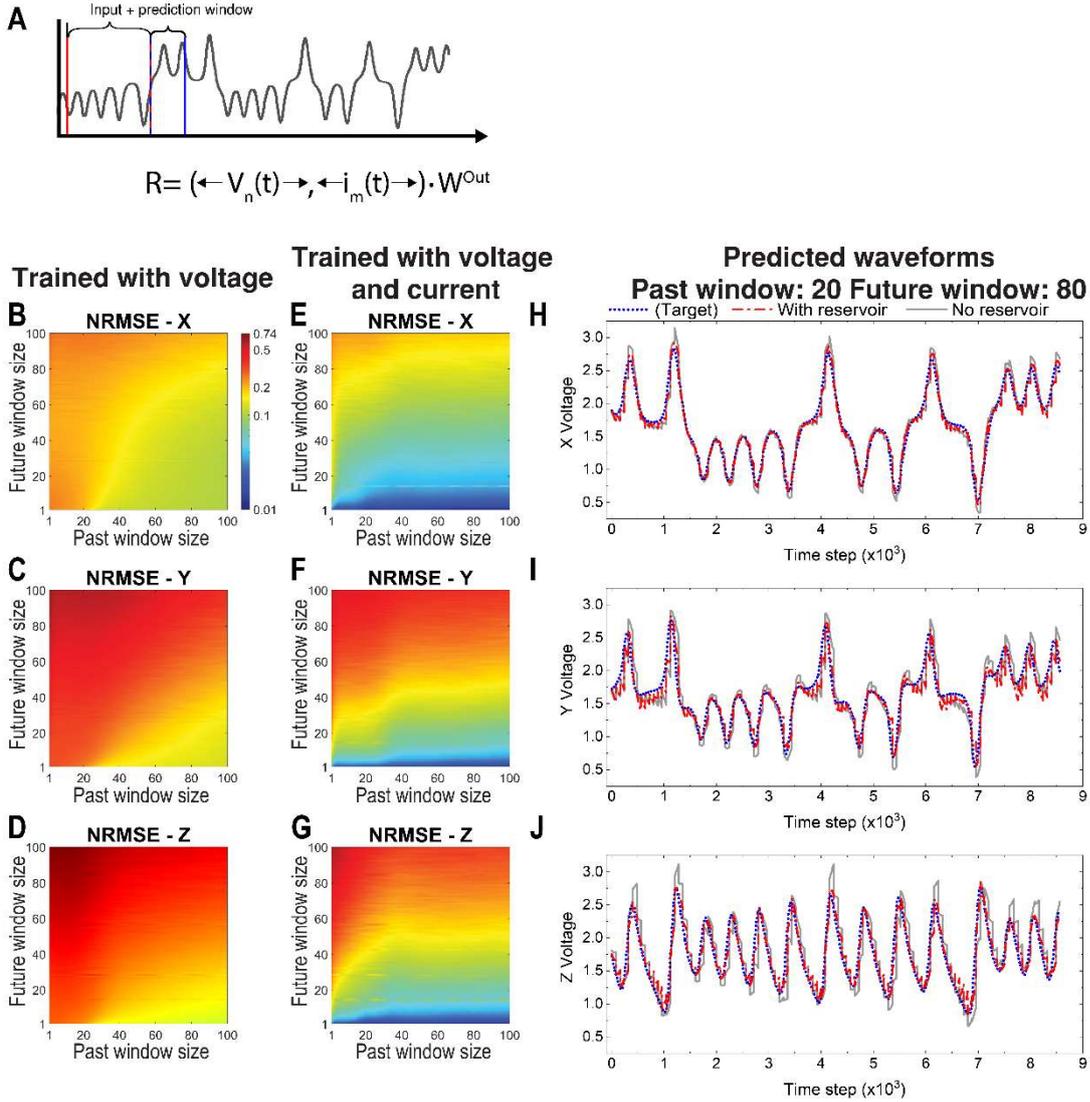

**Figure 6.** A): Schematic illustration of the RC prediction task scheme: a concatenated input row vector, $\{V_n(t), i_m(t)\}$ made from voltage $V_n(t)$ and current $i_m(t)$ data from the input window (red vertical lines), is multiplied with the weight matrix $W^{out}$ to obtain voltage values, $R_{Predicted}$, for the prediction window (blue vertical lines). B)-G): Error values (NRMSE) for prediction of Lorenz-63 time series in form of voltages $V_X(t)$, $V_Y(t)$ and $V_Z(t)$ [top to bottom rows] as calculated using the voltage only as the input, i.e. no reservoir (B) – (D) and using voltage and current inputs, i.e. with the reservoir (E) – (G). H) – J): Comparison between the target voltage waveforms (blue short dash line) $V_X(t)$ (H), $V_Y(t)$ (I) and $V_Z(t)$ (J) and the ones predicted with (red dash line) and without (solid grey line) the reservoir for the past and future windows of 20 and 80 time steps, respectively.

Figure 6B - J shows the outcome of those predictions for different past and future window sizes for all three coordinates. In order to test whether our reservoir provides any improvement, we compare our prediction results based on voltage and current data (that is, with the reservoir) to the same computations performed based on only the voltages (that is, without the reservoir). Our calculations confirm our hypothesis: the corresponding normalized root mean square error (NRMSE) values for predictions without the reservoir (Figure 6B-D) are substantially higher than those with the reservoir (Figure 6E-G), see Figure S3 (A-C) for the comparison plots for the respective error datasets. The corresponding comparison between the prediction curves obtained with and without the reservoir shows clearly the worse performance in case of no reservoir in all three coordinates (Figure 6H-J).

Coming back to the question of the optimal voltage waveform conversion parameters, ($V_0$, $V_{Range}$), this can be described as an example of a physics-aware training.[54] Choosing different voltage amplitudes $V_{Range}$ and offset $V_0$ is aimed at harvesting different non-linearity and short-term memory behavior of the device, and one can think of improving the internal reservoir parameters by obtaining a better fit between the pre-existing physical behavior (for example, its *I-V* response or resistive switching) and the requirements of the given computational problem. We investigated and compared the same task performance with two more combinations ($V_0$, $V_{Range}$) as (1.5 V, 2.5V) and (1.5 V, 3.5V), see Figure S2 for specific NRMSE data and Figure S3 for NRMSE overall comparison. It showed that the combination ($V_0$, $V_{Range}$) = (0.5 V, 2.5V) reported in Figure 6 does indeed yield the lowest NRMSE values among all three voltage conversion parameters.

## 2.6. Lorenz-63 system reconstruction

Beyond predicting future points of the voltage waveform, we also demonstrate a similar computational task, waveform reconstruction: since the original differential equations link all three coordinates together, it should be possible to train our readout to reconstruct two of the three waveforms, for example $V_Y(t)$ and $V_Z(t)$, using only the data for the other remaining waveform, $V_X(t)$, for the same time interval (Figure 7A). We used the same voltage waveform conversion as the one shown in Figure 6, ($V_{DC}$, $V_{AC}$) = (0.5 V, 2.5V). Figure 7B and 7C show reconstruction of 50 time steps in $V_Y(t)$ and $V_Z(t)$ waveforms with 300 time steps in $V_X(t)$ using a time window shifting with each iteration. Again, we compared the network performance with and without the reservoir and found a much better reconstruction performance when the reservoir was in use. We also successfully performed reconstruction of $V_X(t)$ and $V_Z(t)$ from $V_Y(t)$, see Figure S4 A and B. The reconstruction of $V_X(t)$ and $V_Y(t)$ from $V_Z(t)$ failed, both with and without reservoir, see Figure S4 C and D. However, this is in agreement with the Lorenz-63 system properties: its equations (1-3) are symmetric with respect to (*X*, *Y*, *Z*) → (-*X*, -*Y*, *Z*) transformation and hence there are more than one solution to reconstruction based on $Z(t)$ alone, which causes it to fail as a consequence and in agreement with the previous simulation-based results.[55]

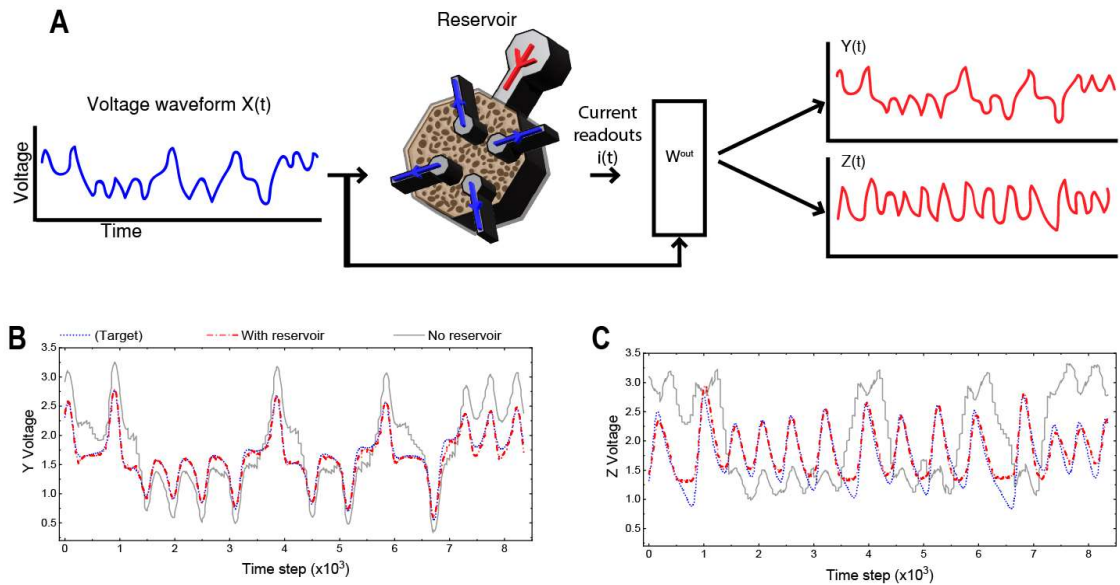

**Figure 7.** (A): Schematic process of reconstruction computations: the input waveform $V_X(t)$ is supplied as voltage to the physical reservoir and the current readouts $i_{X1,2,3}(t)$ are used to reconstruct $V_Y(t)$ and $V_Z(t)$. (B, C): A comparison between the target (short dotted blue line) and computed times series $V_Y(t)$ (B) and $V_Z(t)$ (C) reconstructed without (grey solid line) and with (dashed and dotted red line) reservoir for 50 time step window from 300 time steps window $V_X(t)$, repeated by shifting the window.

Reconstruction of $V_X(t)$ and $V_Z(t)$ from $V_Y(t)$ produced unique results (Figure S4 A, B) since $V_X(t)$ reconstructed from $V_Y(t)$ was the only instance where the results of training from only the original voltage waveform (without the reservoir) produced similar and, in some instances, better results than training with voltage and current, i.e. with the reservoir (Figure S4 A). As $Y(t)$ is a non-linear transformation of $X(t)$, see Equation (1), $X(t)$ should be easier to reconstruct from $Y(t)$ as opposed to $Y(t)$ from $X(t)$, Equation (2), see Figure 7B. To prove this hypothesis, we introduced an additional non-linearity into the system by testing reconstruction of all elements squared, such that $V^2_X(t)$ and $V^2_Z(t)$ were reconstructed from $V^2_Y(t)$. Indeed, as shown in Figure S5, the reconstruction with voltage and current (i.e. with the reservoir) performs better than the one with voltage.

Using the previously reported Lyapunov exponent of 0.906[56] for the given Lorenz system with $s = 10$, $b = 8/3$ and $r = 28$, we calculated the equivalent Lyapunov time for our converted waveforms as being equal to 810 time steps (289 μs in real-world time or an equivalent equation time of 317.11). As such, our best prediction is only valid up to 0.123 Lyapunov times (35.6 µs or an equation time of 39.06). As seen in Figure 6 E - G and Figure S3, the prediction accuracy for larger future window sizes is improving after about 40 time steps in the past window size. That corresponds to about 14 ms in the real time or 0.05 in Lyapunov time and is likely the result of the trade-off between the intrinsic short-term memory of our system and the dynamics of the Lorenz-63 equations. Implementing reservoirs with different thicknesses and with different lateral sizes of top electrodes or varying the porosity should allow to obtain a range of the characteristic short-term memory time constant values and *I-V* curve parameters and thus to adjust the reservoir's performance even further with respect to different desirable output signals. Reversing the roles of bottom and top electrodes, as input and output channels, respectively, could be another interesting future scenario. That is, top electrodes would serve as multiple input channels aimed to sum up several signals into a single (or more than one, in principle) output channel designated at the bottom electrode. The nanoporous oxide structure of the physical reservoir would then enable the mixing of the input signals with random weights rather than splitting the input as in the current scenario.

Our best NRMSE values for Lorenz time series prediction were $1.2 \times 10^{-2}$, $1.5 \times 10^{-2}$ and $1.4 \times 10^{-2}$ for $V_X(t)$, $V_Y(t)$ and $V_Z(t)$, respectively, with a prediction window of 1 time step and a past window of 100 time steps. A comparable work by Yang et al.[35] reported NRMSE values of $1.3 \times 10^{-4}$, $3.1 \times 10^{-4}$ and $2.7 \times 10^{-5}$ for the *X*, *Y* and *Z* components, respectively, for the same predictions window of 1 time step and a past window of 500 time steps, however, they used a 9-output reservoir (vs. 3-output in our case) which was only numerically modelled and not experimentally realized. This shows the potential for improving prediction capability of our devices by increasing the number of input (and / or output) channels, which can be done relatively easily by scaling up the lithography design. Also, our physical system performed satisfactory even for more time steps in the prediction window whilst dealing with fewer outputs, thus minimizing computation time (less iterations) and improving efficiency.

Our system fits into the concept of the so-called in materia computing where the materials system itself is used as a parallel computational medium, e.g. in the form of a network of nanowires[57–67] with physical nodes directly placed within the material, in contrast to other, more conventional approaches of employing a single or multiple separate memristor devices which are similar to each other. In materia computing allows to scale up relatively easily the number of input / output channels of the reservoir with minimal hardware costs. Previous realizations were either lacking in demonstrations of challenging computational tests such as prediction of chaotic time series, for example, or were based on non-oxide materials such as iodides, sulfites and organic materials.[57–62,68] On the other hand, memristor device-based physical RC approaches struggle to deal with issues of connectivity and separability when a higher number of similar devices are to be interlinked and employed.[57] Here we unite both concepts of oxide memristor-based physical RC with the one of in materia computing and show that it is indeed possible to achieve computational performance which is at least comparable to other oxide memristor-based systems whilst maintaining the benefits of in materia computing.

## 3. Conclusion

We demonstrated an all-oxide reservoir system designed intentionally with a random inhomogeneity in the resistive switching oxide layer where incoming voltage input is mixed and redirected from the common bottom electrode into different output top electrodes, and multiple channel currents are recorded in parallel. This allows us to use this physical reservoir for even challenging tasks of prediction and reconstruction of chaotic Lorenz-63 times series. We have also demonstrated the other promising capabilities of our physical reservoir such as image recognition, solving the XOR task and waveform reconstruction. The latter can be used to generate synthetic data in the case of data scarcity, for example, when missing one of the multiple input data channels. Future development should also encompass combinations of multiple reservoirs in form of either deep[69,70] or parallel RC[71,72] to allow for increasing the complexity of the input data.

## 4. Experimental Section

*Device fabrication:* The devices were fabricated by magnetron sputtering and UV lithography on Si/SiO$_2$ substrates with a 200 nm SiO$_2$ top layer. The complete device stack was comprised, from the bottom to the top, of 20 nm nanoporous platinum (np-Pt), 80 nm nitrogen-doped NbO$_x$ (SEM imaged device was 55 nm), 3 nm Ti, and 120 nm Pt.

The np-Pt bottom electrodes were produced using a method adapted from Jung *et al*. [73]. 20 nm Pt was deposited in an Ar:O$_2$ mixture with flow rates of 9.0 sccm and 6.0 sccm, respectively, by DC magnetron sputtering (100W) at 5.9 mTorr and a deposition rate of 0.16 nm/s. The substrate was at room temperature. This resulting PtO$_x$ film was then annealed for 1 hour at 600 °C under high vacuum conditions with heating and cooling performed at a ramp rate of 20 °C / minute. Subsequently, photoresist coating, UV photolithography, argon ion milling, and a lift-off process were used to fabricate bottom electrodes with a width of 50 μm. See supplementary information for more details on the spin-coating steps. Before the resistive layer was deposited, the substrate was coated with two photoresists, to encourage an undercut during the patterning process. The resistive nitrogen-doped $NbO_x$ layer, 80 nm (53 nm as in Figure 2), was deposited via RF magnetron sputtering (200W) in an $Ar/O_2/N_2$ atmosphere with the respective flow rates of 5.0, 11.7, and 0.7 sccm at 3.1-3.7 mTorr and a deposition rate of 0.10 nm/s. The inclusion of $N_2$ during this step significantly increases the resistivity of the layer compared to those deposited in an $Ar/O_2$ atmosphere alone[49]. The nitrogen-doped $NbO_x$ layer covered much of the bottom electrode except for the electrode contact paths and pads (Figure 2A). The patterning of the $NbO_x$ layer made use of two photoresists to develop an undercut in order to reduce the sharpness of the edge and provide a more rounded edge for the next layer to deposit on. After a final photolithography step, the top electrodes were deposited by DC magnetron sputtering (100W), comprising of 3 nm Ti adhesion layer and 120 nm Pt layer. These electrodes were 10 μm in width, with some overlap from the trace to the contact pad. The Ti (Pt) layer was deposited with an Ar flow rate of 5.0 (5.0) sccm, a growth pressure of 2.7 (2.6) mTorr and a deposition rate of 0.06 (0.28) nm/s.

Contact with the electrodes were made by 0.7 um W probe tips. *I-V* characteristics and pulsed electrical measurements were conducted using a Keithley 4200A-SCS. DC current-voltage sweeps were performed at a rate of 0.75 V/s.

*Readout training for XOR task:* The SLP was set up in MATLAB R2024a using the built-in function for networks, called 'perceptron'. The current readout values ($x$) were normalized by calculating Z-scores defined as:

$$Z = \frac{(x - \bar{x})}{\sigma},$$

where $\bar{x}$ is the mean value and $\sigma$ is the standard deviation. By including virtual nodes, we measured a total of 7 data sets which were split into 3 for training and 4 for testing. Including the reservoir outputs increases the size of the input data vector for the perceptron function from two data points to five: the two original input values (0 or 1) and the three reservoir readout values.

*Readout training for number recognition:* The output layer training was implemented in MATLAB R2024a using the function fmincg.m created by Carl E. Rasmussen.

Each input vector $U$ containing five current readout values produces an output vector $R$ of ten probabilities for the hypothesis that this specific input $U$ corresponds to one of the digits "0" to "9":

$$R(U) = g(U \cdot W^{out} + B)$$

where, $W^{out}$ is the weight matrix, $B$ is the bias vector of the same length as $R$ and $g$ is the logistic function:

$$g(z) = \frac{1}{1+e^{-z}}$$

Weight matrix $W^{out}$ and bias vector $B$ are created by minimizing $k$-th column and $k$-th element in $W^{out}$ and in $B$, together and for each $k$ corresponding to one of the digits "0" to "9", using the following cost function [74] with training data:

$$J_k = \frac{1}{L}\sum_{l=1}^{L}\left[-H_{l,k}\log\left\{g\left(\sum_{j=1}^{5}U_{l,j}^{\text{Train}}\cdot W_{j,k}^{out}+B_k\right)\right\}\right.$$

$$\left.-(1-H_{l,k})\log\left\{1-g\left(\sum_{j=1}^{5}U_{l,j}^{\text{Train}}\cdot W_{j,k}^{out}+B_k\right)\right\}\right]+\frac{\lambda}{2L}\sum_{j=1}^{5}(W_{j,k}^{out})^2$$

where $l$ runs up to $L$ rows (i.e. total number of data samples) within the training input signal matrix $U_{l,j}^{\text{Train}}$. Each minimization procedure for a $k$-th column ($k$-th element if in $B$) uses a target probability vector $H_k$ which contains $H_{l,k}=1$ if the $l$-th input row in $U^{\text{Train}}$ does indeed represent the desired $k$-th digit and $H_{l,k}=0$ if otherwise. The last term represents L2 or Tikhonov regularization [75]. In our case $\lambda = 10^{-5}$. The cost function is minimized using the following gradients

$$\frac{\partial J_k}{\partial W_{n,k}^{out}} = \frac{1}{L}\sum_{l=1}^{L}\left[g\left(\sum_{j=1}^{5}U_{l,j}^{\text{Train}}\cdot W_{j,k}^{out}+B_k\right)-H_{l,k}\right]U_{l,n}^{\text{Train}}+\frac{\lambda}{L}W_{n,k}^{out}$$

$$\frac{\partial J_k}{\partial B_k} = \frac{1}{L}\sum_{l=1}^{L}\left[g\left(\sum_{j=1}^{5}U_{l,j}^{\text{Train}}\cdot W_{j,k}^{out}+B_k\right)-H_{l,k}\right]$$

*Waveform conversion*

$$V(t) = V_0 + V_{\text{Range}}\frac{f(t)-f_{\min}}{f_{\max}-f_{\min}}$$

$V_0$ and $V_{\text{Range}}$ are the voltage bias and range respectively. $f(t)$ is the Lorenz time series, $X(t)$, $Y(t)$ or $Z(t)$. $f_{\min}$ and $f_{\max}$ denote the minimal and maximal values for the given time range. For computations, the voltage waveforms $V(t)$ were normalized to $V'(t)$: $V'(t) = \frac{V(t)-V_{\min}}{V_{\max}-V_{\min}}$. $V_{\min}$ and $V_{\max}$ denote the minimal and maximal voltage values for the given time range.

*Ridge regression and output neural network layer:* We denote previous and future/reconstruction window sizes are $N^{past}$ and $N^{out}$, respectively, $N$ is the total dataset size, $a$ is the training ratio ($a = 0.7$ in our case), $g$ and $d$ is the number of input and output variables, respectively.

Actual waveform computations were performed in an open-loop approach by sequentially shifting the output data window $N^{out}$ related to output $R$ across the data set with the corresponding past window $N^{past}$ related to the input $U$ using the equation:

$$R = U \cdot W^{out}$$

where $R$, $U$ and $W^{out}$ are output and input row vectors, and the weight matrix, respectively, of the following matrix dimensions represented as

{number of rows × number of columns}:

$U = \{1 \times \gamma N^{past}\}$ and $R = \{1 \times \delta N^{out}\}$

$$W^{out} = \{\gamma N^{past} \times \delta N^{out}\}$$

For predictions: $\gamma = 3$ with no reservoir input (3 voltages), $\gamma = 12$ with the reservoir input (i.e. 3 voltages and 9 current outputs); $\delta = 3$ for all cases:

$$U = (V_X^{past}, V_Y^{past}, V_Z^{past}) \text{ or } U = (V_X^{past}, V_Y^{past}, V_Z^{past}, i_{X1,2,3}^{past}, i_{Y1,2,3}^{past}, i_{Z1,2,3}^{past})$$

$$R = (V_X^{out}, V_Y^{out}, V_Z^{out})$$

$V_X^{past}$, $V_X^{out}$ and $i_{X1,2,3}^{past}$ denote input and output voltages and output currents (3 outputs) datasets of the lengths $N^{past}$ and $N^{out}$ each, respectively, for the $X(t)$ waveform, etc.

For reconstruction, we used $\gamma = 1$ with no reservoir, $\gamma = 4$ (i.e. 1 voltage and 3 currents) with the reservoir; $\delta = 2$ for all cases.

The weight matrix $W^{out}$ was calculated during the training based on 70% of the whole waveform using the following ridge regression equation [11]:

$$W^{out} = (U_{Train}^T U_{Train} + \lambda I)^{-1} U_{Train}^T R_{Train}$$

under assumption $R_{Train} = U_{Train} \cdot W^{out}$, where $U_{Train}$ is the matrix containing all the training data sets of the corresponding input voltage and output currents, $U_{Train}^T$ is the transpose of $U_{Train}$, $\lambda$ is the L2 regularization parameter, $I$ is a square identity matrix of the same size as number of columns in $U_{Train}$. $R_{Train}$ is the matrix of all target voltage values used for training. Each row in $U_{Train}$ and in $R_{Train}$ was composed the same way as in $U$ and $R$, respectively. Each prediction window is calculated independently. The past and output window sizes, $N_{past}$ and $N_{out}$, were varied from 1 to 100 and the L2 parameter was set to $\lambda = 1.5$.

The matrix dimensions were

$$U_{\text{Train}} = \left\{ \left[ \frac{\alpha N}{N_{\text{out}}} \right] \times \gamma N_{\text{past}} \right\}$$

$$R_{\text{Train}} = \left\{ \left[ \frac{\alpha N}{N_{\text{out}}} \right] \times \delta N_{\text{out}} \right\}$$

Matrices $U_{\text{Train}}$ and $R_{\text{Train}}$ were filled by moving the window $N_{\text{out}}$ until a rounded-up number of rows $\left[ \frac{\alpha N}{N_{\text{out}}} \right]$ was completed. For predictions, the window $N_{\text{out}}$ was started after the first window, $N_{\text{past}}$, which means the last row of training data was allowed to be taken beyond the $aN$ training limit by at least $N_{\text{past}}$ and as much as $N_{\text{past}} + N_{\text{out}}$ (~1-2% of $N$) For reconstruction, the window $N_{\text{out}}$ was taken as to complete at the end of the window $N_{\text{past}}$ so the overshoot beyond the $aN$ training limit was at least $N_{\text{past}} - N_{\text{out}}$ and as much as $N_{\text{past}}$ (~3.5% of $N$).

*Normalized root mean square error:* We use the normalized root mean square error (NRMSE), normalized with respect to the variance and calculated as follows:

$$\text{NRMSE} = \frac{\sqrt{\frac{1}{n} \sum_{i=1}^{n} (V_i - V_i^{\text{Target}})^2}}{\sqrt{\frac{1}{n-1} \sum_{i=1}^{n} (V_i - \langle V^{\text{Target}} \rangle)^2}}$$

where $n$ is the total number of data points, $V_i^{\text{Target}}$ is the actual voltage value, $V_i$ is the predicted votage value and $\langle V^{\text{Target}} \rangle$ is the mean of the target voltage.

**Acknowledgements**

The authors acknowledge use of the facilities and the assistance of Dr Zhaoxia Zhou, Dr Stuart Robertson and Sam Davis in the Loughborough Materials Characterization Centre and for access to the NOVA 600 Nanolab Ga DualBeam and the JEOL JSM 7800F FEGSEM. This work is supported by the UK Engineering and Physical Sciences Research Council (EPSRC), Grant No. EP/T027479/1 and studentship 2764909.

**Data Availability Statement**

All data are available in the main text or the supplementary materials. Additional data related to this paper may be requested from the authors.

Supporting Information

**Scalable platform enabling reservoir computing with nanoporous oxide memristors for image recognition and time series prediction**

*Joshua Donald, Ben A. Johnson, Amir Mehrnejat, Alex Gabbitas, Arthur G. T. Coveney, Alexander Balanov, Sergey Savel'ev, Pavel Borisov\**

**Supplementary Text**
Photoresist steps

For the np-Pt bottom electrode: coating with AZ3007 photoresist (Merck Performance Materials GmbH) in inverted patterns, so that ion milling (etched for 1.5 minutes at $2.4\times10^{-4}$ Torr, with an accelerating potential of 99V and current 1.5 mA to produce a beam of 497V and 37.6 mA) can be used to remove the unwanted material, annealed for 1 min at 110 °C and then patterned.

For $NbO_x$ layer: coating with LOR3A photoresist (Kayaku), annealing for 3 minutes at 180 °C, then coating with AZ3007 and annealing for 1 minute at 110 °C. This pattern was not inverted.

For Pt top electrodes: Same parameters as the bottom electrode, but the pattern was not inverted.

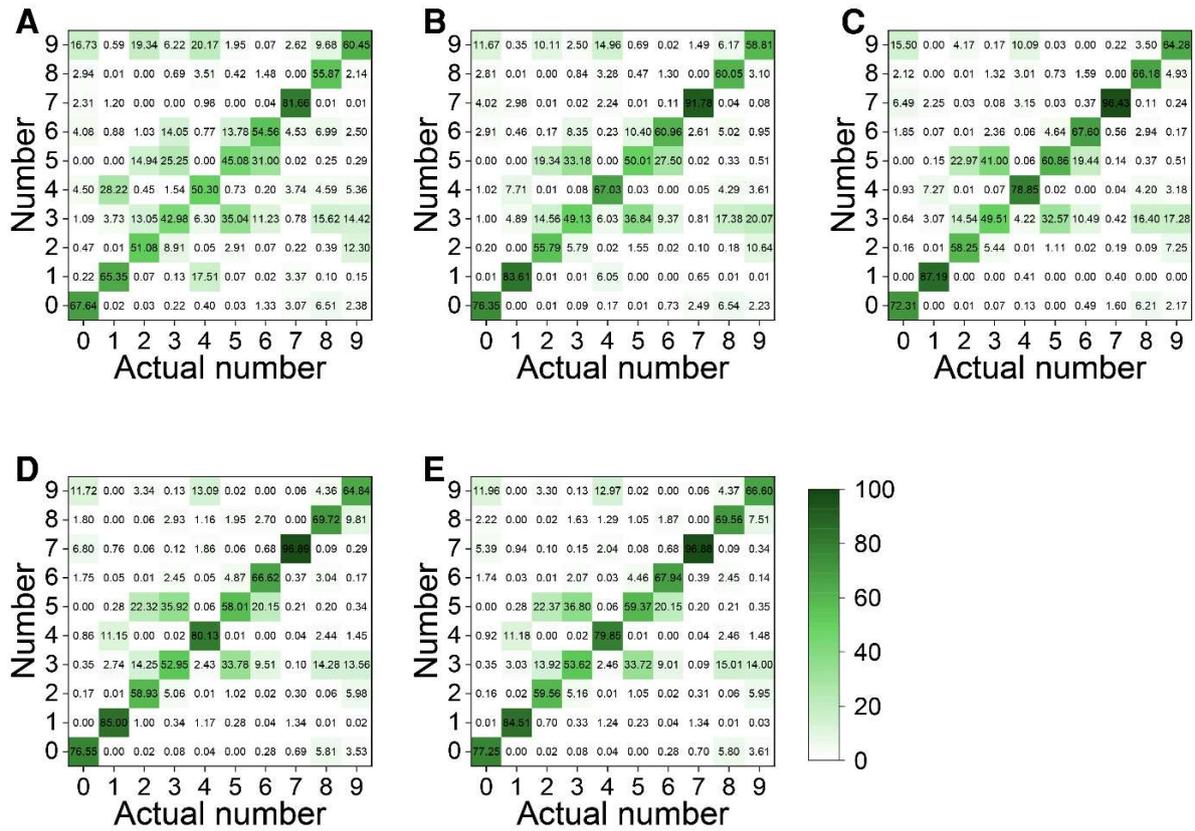

**Figure S1.** Average normalized prediction scores provided as output when performing the number recognition task. after 10 (A), 14 (B), 50 (C), 100 (D) and 200 (E) training iterations, respectively. The horizontal axis represents the actual digit image of which is to be recognized. The vertical axis denotes the average probability for each possible digit. The highest prediction score is taken as the predicted value.

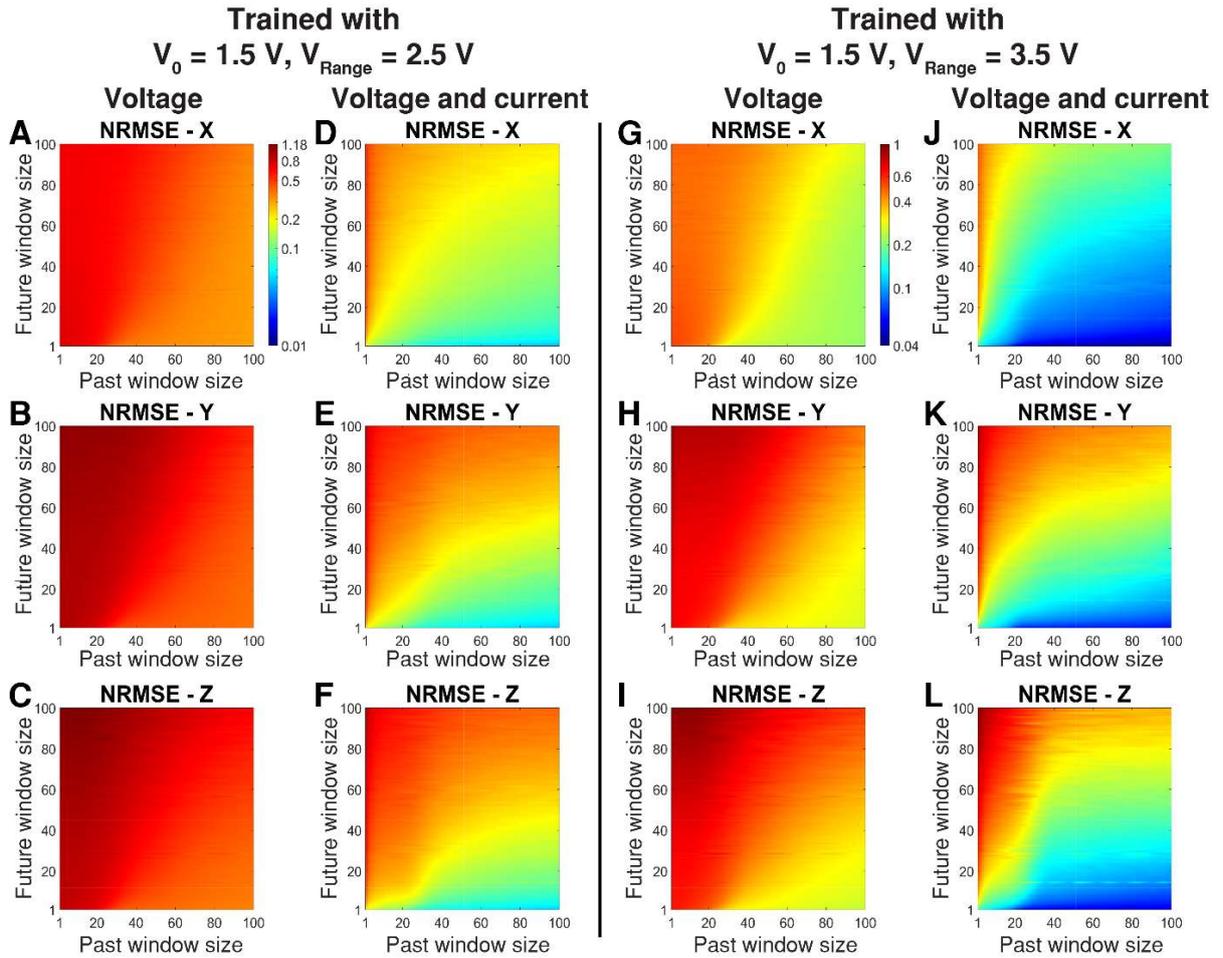

**Figure S2.** Alternative voltage conversion parameters ($V_0$, $V_{\text{Range}}$): NRMSE values for prediction of Lorenz-63 time series in form of voltages $V_X(t)$, $V_Y(t)$ and $V_Z(t)$ [top to bottom rows] as calculated using the voltage only as the input, i.e. no reservoir (A-C) and (G - I) and using voltage and current inputs, i.e. with the reservoir (D-F) and (J-L). Figure A-F (left) and G-L (right) were obtained for ($V_0$, $V_{\text{Range}}$) = (1.5V, 2.5V) and (1.5V, 3.5V), respectively. Blue values denote a low NRMSE, indicating a better prediction. Red values denote a larger NRMSE, indicating a worse prediction.

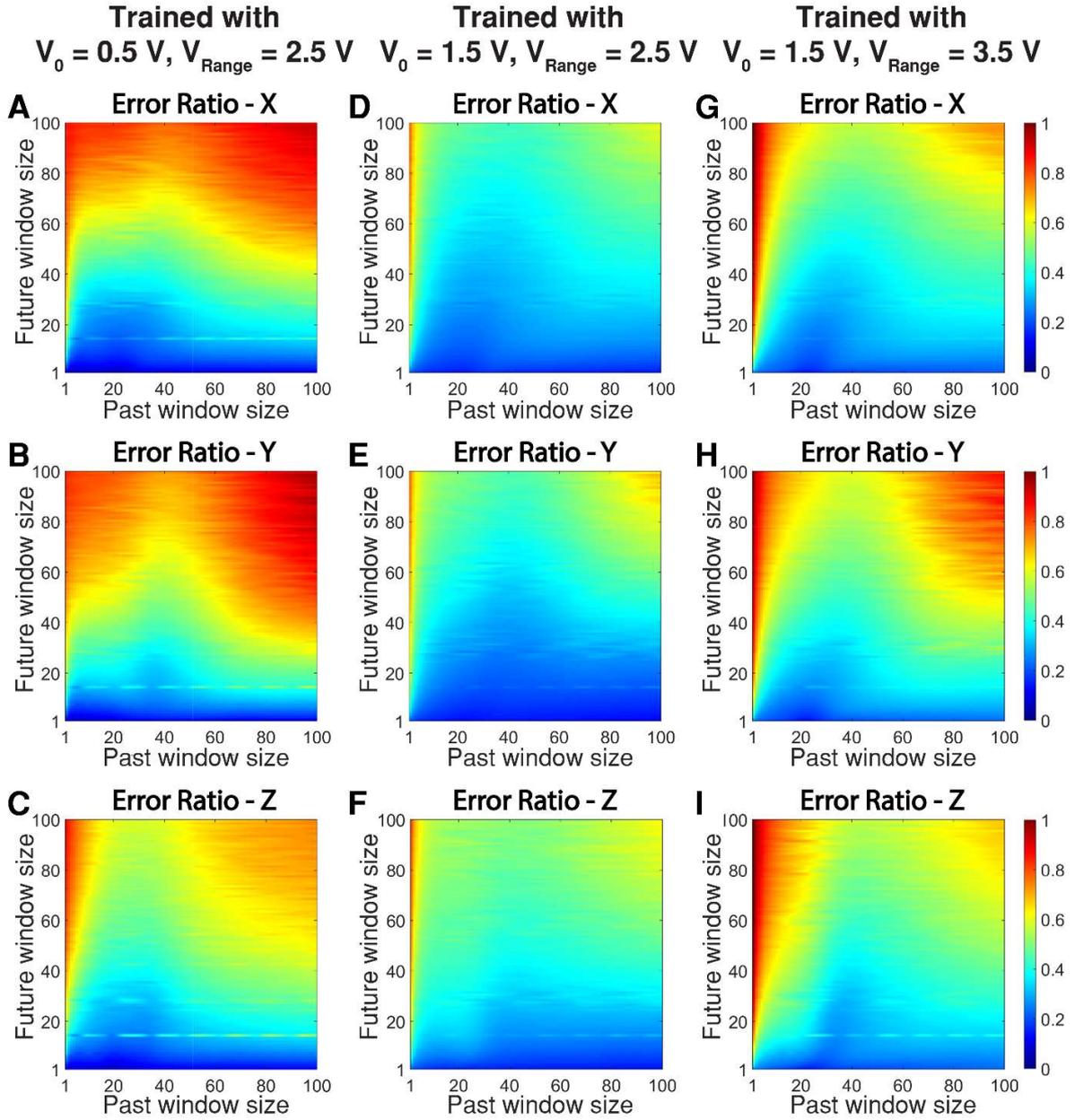

**Figure S3.** Comparison of voltage conversion parameters ($V_0$, $V_{Range}$): ratio of NRMSE values for prediction of Lorenz-63 time series in form of voltages $V_X(t)$, $V_Y(t)$ and $V_Z(t)$ [top to bottom rows]. The ratio is taken as NRMSE for the case voltage and current inputs (with reservoir) divided by NRMSE for the case of voltage only input (no reservoir); the lower the number, the greater the improvement. Three voltage conversion parameters are considered: (A)-(C): $V_0 = 0.5V$, $V_{Range}=2.5V$. (D)-(F): $V_0 = 1.5V$, $V_{Range}=2.5V$. (G)-(I): $V_0 = 1.5V$, $V_{Range}=3.5V$.

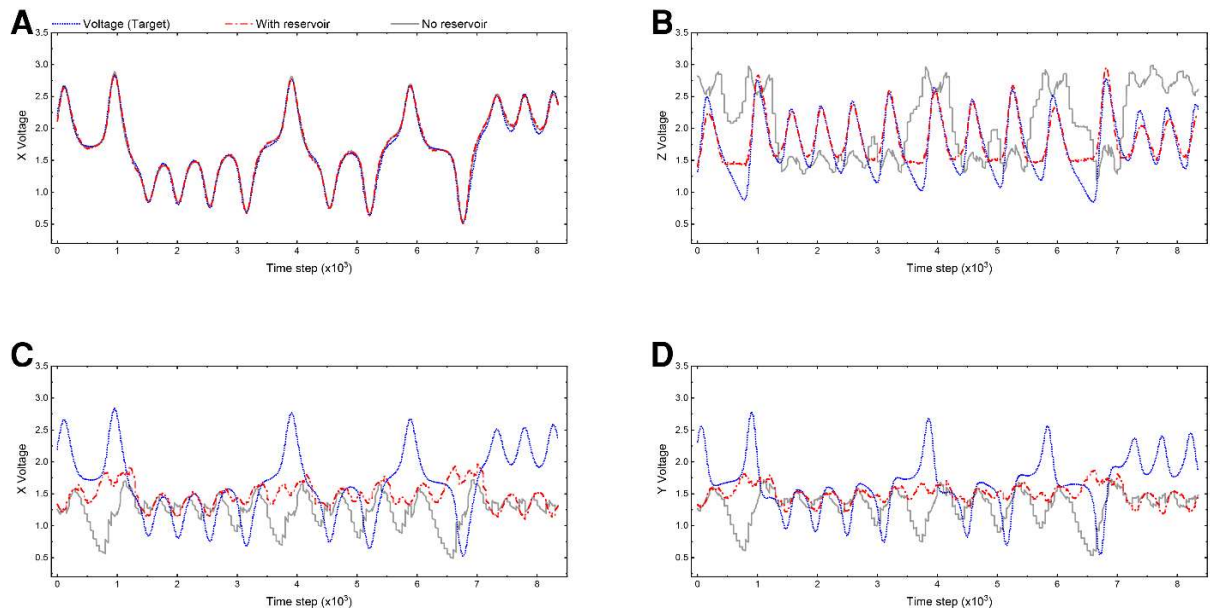

**Figure S4.** Reconstruction task: a comparison between the target (short dotted blue line) and computed times series reconstructed without (grey solid line) and with (dashed-dotted red line) reservoir for 50 time step window from 300 time steps window of $V_Y(t)$ [reconstructed voltages $V_X(t)$ (A) and $V_Z(t)$ (B)] and $V_Z(t)$ [reconstructed voltages $V_X(t)$ (C) and $V_Y(t)$ (D)].

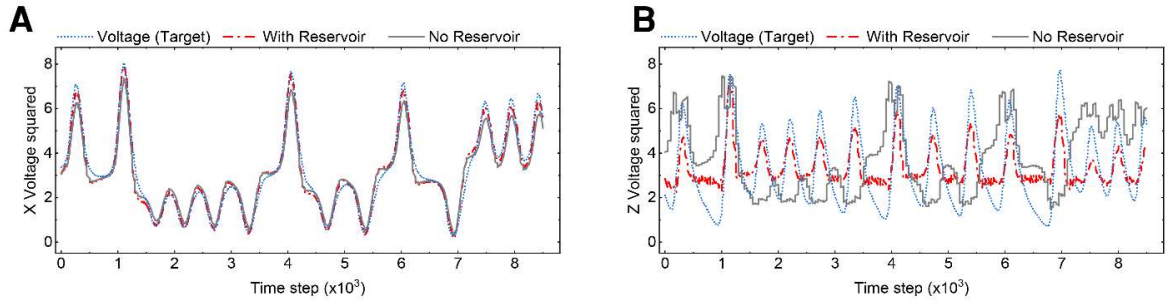

**Figure S5.** Reconstruction task: a comparison between the target (short dotted blue line) and computed times series $V^2{}_X(t)$ (A) and $V^2{}_Z(t)$ (B) reconstructed without (grey solid line) and with (dashed and dotted red line) reservoir for 50 time step window from 300 time steps window of $V^2{}_Y(t)$, repeated by shifting the window.